\documentstyle[12pt]{article}
\begin{document}
\thispagestyle{empty}

\begin{center}
\LARGE \tt \bf{Sachs-Wolfe effect in Spacetimes with Torsion}
\end{center}

\vspace{2.5cm}

\begin{center} {\large L.C. Garcia de Andrade\footnote{Departamento de
F\'{\i}sica Te\'{o}rica - Instituto de F\'{\i}sica - UERJ

Rua S\~{a}o Fco. Xavier 524, Rio de Janeiro, RJ

Maracan\~{a}, CEP:20550-003 , Brasil.

E-mail : GARCIA@DFT.IF.UERJ.BR}}
\end{center}

\vspace{2.0cm}

\begin{abstract}
Cartan torsion contribution to Sachs-Wolfe effect in the inflationary phase of the Universe is discussed.From the COBE data of the microwave anisotropy is possible to compute the spin-density in the Universe as $10^{16}$ mks units.The spin-density fluctuations at the hadron era is shown to coincide with the anisotropy temperature fluctuations.
\end{abstract}

\vspace{1.0cm}

\begin{center}
\large{PACS number(s) : 0420,0450}
\end{center}

\newpage
The Sachs-Wolfe effect \cite{1,2,3} is given by
\begin{equation}
\frac{{\delta}T}{T}=-{\Phi}
\label{1}
\end{equation}
where ${\Phi}$ is the Newtonian gravitational potential.Since there vis no restriction on the origin of the gravitational potential in this expression one may extend this expression to allow for alternative theories of gravity.Thus one is able to compute the anisotropy fluctuations in the CMB temperatures.In this letter we shall be concerned with the computation of the temperature fluctuations in the context of weak field limit approximation of the Einstein-Cartan gravity.In Newtonian approximation the Einstein-Cartan field equation may be expressed as
\begin{equation}
\frac{1}{c^{2}}{\frac{{\partial}^{2}{\Phi}}{{{\partial}t}^{2}}}-{{\nabla}^{2}}{\Phi}=-4{\pi}G({\rho}-G{\sigma}^{2})
\label{2}
\end{equation}
where ${\sigma}^{2}$ is the spin-torsion energy density and ${\rho}$ is the usual matter density.Since the cosmological spacetime used is of the form
\begin{equation}
ds^{2}=dt^{2}-(1-2{\Phi})(dx^{2}+dy^{2}+dz^{2})
\label{3}
\end{equation}
is homogeneous ${{\nabla}^{2}}{\Phi}=0$ equation (\ref{2}) reduces to
\begin{equation}
\ddot{\Phi}=-4{\pi}Gc^{2}({\rho}-G{\sigma}^{2})
\label{4}
\end{equation}
Solution of this expression can be easily obtained obtained by considering the RHS of (\ref{4}) is constant in the average, thus
\begin{equation}
{\Phi}(t)=-4{\pi}Gc^{2}({\rho}-G{\sigma}^{2})t^{2}+Bt+C
\label{5}
\end{equation}
where $B$ and $C$ are integration constants.Therefore spin-torsion effects contribute to the temperature fluctuations through the term
\begin{equation}
{\frac{{\delta}T}{T}}_{torsion}=4{\pi}G^{2}c^{2}{\sigma}^{2}t^{2}
\label{6}
\end{equation}
Since at the inflation era $t=10^{-35}s$ and from COBE data $\frac{{\delta}T}{T}<10^{-5}$ one may obtain an expression for the spin-torsion density at the inflation era as
\begin{equation}
{\sigma}^{2}=\frac{1}{4{\pi}}(G^{-2}c^{-2}t^{-2}){\frac{{\delta}T}{T}}_{torsion}
\label{7}
\end{equation}
which yields ${\sigma}=10^{16}$ mks-units.Notice that this value is much weaker than the value obtained by de Sabbata and Sivaram \cite{4} for the maximum spin-torsion density at the Planck era of the Universe which is ${\sigma}_{Pl}=10^{71}$.In the hadron era \cite{4} for example $G^{2}{\sigma}^{2}=T^{2}$ and one may obtain the classical density fluctuation as 
\begin{equation}
\frac{{\delta}T}{T}=\frac{{\delta}{\sigma}}{\sigma}
\label{8}
\end{equation}
This amazing result shows that the spin-torsion density fluctuation coincides with the matter density fluctuation which allow us to obtain the spin-torsion density fluctuation from the COBE data.Similar results in the Relativistic Einstein-Cartan gravity with dilaton fields has been recently obtained by Palle \cite{5} and myself \cite{6}.
\section*{Acknowledgement}
I am very much indebt to Prof.I.Shapiro and Prof.R.Ramos for helpful discussions on the subject of this paper. Financial support from CNPq. and FAPESP is gratefully acknowledged.

\end{document}